\documentstyle[preprint,aps]{revtex}
\input{psfig}
\newcommand{\bra}[1]{\mbox{$\langle #1|$}}
\newcommand{\ket}[1]{\mbox{$|#1\rangle$}}
\newcommand{\braket}[2]{\mbox{$\langle #1 | #2 \rangle$}}

\begin{document}
\preprint{FNT/T 2002/20, MKPH-T-02-24}

\title{{\bf On the NN-final-state-interaction in the
  $^{16}\mbox{O}(e,e' pp)$ reaction}}
 \author{M.\ Schwamb}
\address{Institut f\"ur Kernphysik,
Johannes Gutenberg-Universit\"at, D-55099 Mainz, Germany}
\author{ S.\ Boffi, C.\ Giusti and  F.\ D.\  Pacati}
\address{Dipartimento di Fisica Nucleare e Teorica
dell'Universit\`a, Pavia and Istituto Nazionale di Fisica Nucleare,
Sezione di Pavia, I-27100 Pavia, Italy}
\maketitle

\begin{abstract}
\noindent
The influence
 of the mutual interaction between the two outgoing nucleons
(NN-FSI) in the $^{16}\mbox{O}(e,e' pp)$ reaction has been investigated.
  Results for various kinematics are discussed.  In general, the effect
 of NN-FSI depends  on kinematics and the chosen final state 
in the excitation spectrum of $^{14}\mbox{C}$. \\

\end{abstract}

\section{Introduction}
\label{intro}

The independent particle shell model, describing a nucleus as a  system 
 of nucleons moving in a mean field, reproduces  many basic features of nuclear
 structure. It is however well-known that the  repulsive components 
of the NN-interaction induce additional short-range correlations (SRC)
 which are beyond a mean field description and whose investigation
 can  provide additional  insight into the nuclear structure. 
 A  powerful tool for the investigation of   SRC are
 electromagnetic  two-nucleon knockout reactions like $(\gamma,NN)$  or
 $(e,e'NN)$, since 
the probability that a real or a virtual photon is absorbed by a pair of 
 nucleons should be a direct measure for the correlations between these
nucleons (for an overview, see \cite{Bof96}). 
 However, this simple picture has to be modified because  additional
 complications have to be taken into account.
 In particular, competing mechanisms like  contributions
 of two-body currents as well as final state interactions (FSI)
 between the two outgoing nucleons and the residual nucleus 
 have to be considered. However, it turned out in previous studies
 \cite{Gui97,Gui98,Gui99} that it is --at least in principle--
 possible to determine specific kinematical situations
 where the reaction cross section is particularly sensitive to SRC.

 Due to the complexity of the subject, several approximations
 have been performed in the past, which restrict the reliability of the
 existing models (consider
 \cite{Bof96,Gui97,Gui98,Gui99,Gui00,Gui01,Ryc97,Ryc98a,Ryc98b}
 and the references
 therein) with respect to the interpretation of the existing 
 experimental data.  In this context, one crucial assumption is the
 fact that the mutual interaction between the two outgoing nucleons,
 denoted as NN-FSI, can be neglected. At the moment no reliable
 estimate of this approximation exists  for two-nucleon
 knockout on finite nuclei.  Indeed, recent calculations on nuclear
 matter \cite{Kno00} clearly indicate that NN-FSI are non-negligible
 even if the two detected nucleons are ejected back to back in the
 so-called superparallel kinematics, where NN-FSI are expected to be
 minimal. However, a study in nuclear matter does not provide results
 for cross sections which can directly be compared with experimental
 data produced for a specific target nucleus.

A consistent treatment of FSI would require in general 
 a genuine three-body
 approach for the mutual interaction of the two protons and the 
residual nucleus (see fig.\  \ref{fig_mechanism}).
 Presumably due to the enormous computational challenges,
 this has never been tackled in the past. Before starting such an 
ambitious project, it may be recommendable  to estimate first within an
 approximative, but more feasible  treatment
 the qualitative role of   NN-FSI. This has been done in the
present work  using as underlying framework the unfactorized approach
 for two-nucleon knockout on complex, but finite nuclei
presented in \cite{Gui97}.  If NN-FSI in these  studies turned
out to be small,  a  complete three-body calculation would 
not be necessary and the previous treatment of neglecting
NN-FSI completely could  be justified.

The paper is organized as follows. 
The theoretical framework
and the adopted approximations are outlined in sect. \ref{mod}.  Numerical
results for some selected kinematical situations are presented in
sect. \ref{res}, where also some perspectives of possible improvements 
 and future developments are given.

\section{The model}
\label{mod}
The cross section for electromagnetic two-proton-knockout is given in
general by the square of the scalar product of the relativistic
electron current $j^{\mu}$ and of the nuclear current $J^{\mu}$, where
the latter is given by the Fourier transform of the transition matrix
element of the charge-current density operator between initial and final
 nuclear states
\begin{equation}\label{eq1}
J^{\mu}(\vec{q}\,) = \int \bra{\Psi_f} \hat{J}^{\mu}(r) \ket{\Psi_i}
e^{i \vec{q} \cdot \vec{r}} \,\,.
\end{equation}

 Concerning the nuclear current
$\hat{J}^{\mu}(r)$ and the initial state $\ket{\Psi_i}$ of the two emitted
protons, the general framework described  in  \cite{Gui97}
has been adopted without any modification. Thus, the
nuclear current operator $\hat{J}^{\mu}(r)$ is the sum of 
 a one- and a two-body part. The one-body part
 consists of  the usual  charge operator and the  
convection and spin current. In the two-body part the nonrelativistic 
 pionic seagull and flight meson-exchange current
 do not contribute in two-proton emission, so that only intermediate
 $\Delta$ isobar excitation has to be considered \cite{WiA97}.

For the $^{16}\mbox{O}(e,e' pp)$  reaction, the initial state
 $\ket{\Psi_i}$  is taken from the calculation of the 
two-proton spectral function of $^{16}\mbox{O}$ in
 \cite{Geu96}, where long-range
and short-range  correlations are consistently taken into account.
 The latter are included in the radial wave functions  of relative motion 
 through defect functions, which were obtained by solving the Bethe-Goldstone
 equation for  $^{16}\mbox{O}$ in momentum space.

\begin{figure}
\centerline{\psfig{figure=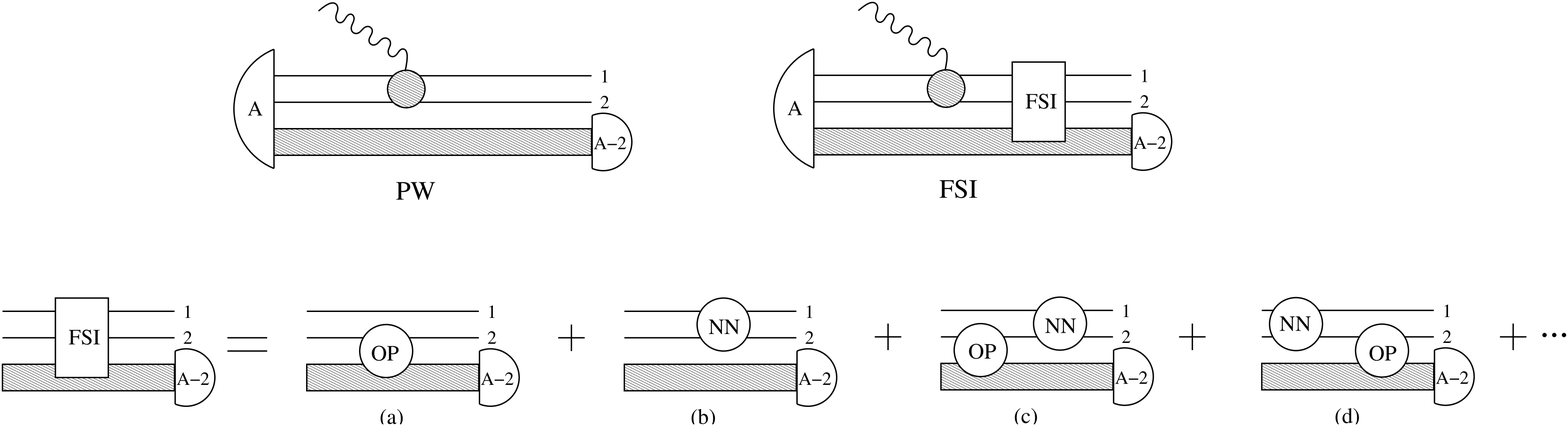,width=16cm,angle=0}}
\vspace{0.5cm}
\caption{The relevant diagrams for electromagnetic 
 two-nucleon knockout on a complex nucleus
A. The two diagrams on top depict the plane-wave approximation 
 (PW)  and  the distortion of the two outgoing proton wave functions
by final state interactions (FSI).  Below, the relevant mechanisms of 
FSI are depicted in detail, where the open circle denotes either a 
 nucleon-nucleus interaction given by a phenomenological optical potential
(OP) or the mutual interaction between the two outgoing nucleons (NN). 
Diagrams which are given by an interchange of nucleon 1 and 2 are not depicted.
}
\label{fig_mechanism}
\end{figure}

As has already been outlined in the introduction, 
several  approximations have 
 been used in the past concerning the final state $\ket{\Psi_f}$.
In the simplest approach  any interaction between the two 
 protons and the residual nucleus is neglected and a plane-wave approximation 
 (PW) is assumed for the two outgoing protons
 (see fig.\ \ref{fig_mechanism}).
 In the more sophisticated
 approach of \cite{Gui97},  the interaction between each of the outgoing 
 protons and the residual nucleus is considered by using a complex 
 phenomenological optical potential $V^{OP}$ for nucleon-nucleus scattering  
 which contains a central, a Coulomb and a spin-orbit term \cite{Nad81} (see 
 diagram (a) in fig.\ \ref{fig_mechanism}).
 Under the simplifying assumption of an infinite heavy residual nucleus,
 the corresponding final state  can be expressed  as   the 
 {\it product} of two uncoupled single-particle distorted wave functions 
 \braket{\vec{r}_i}{\phi^{OP}(\vec{p}^{\,\,0}_i)} ($i = 1,2$). The latter
 are given by  the solution of the corresponding  Schr\"o\-dinger equation
\begin{equation}\label{op-equation}
\left( H_0(i) + V^{OP}(i) \right) \ket{\phi^{OP}(\vec{p}^{\,\,0}_i)} = E_i 
\ket{\phi^{OP}(\vec{p}^{\,\,0}_i)},
\end{equation}
 with $H_0(i)$ denoting the kinetic energy operator 
 and $\vec{p}^{\,\,0}_i$ the asymptotic free momentum of the outgoing 
 proton $i$, with kinetic energy $E_i$ in the used laboratory frame.  In 
practice, the finite mass $m_{^{14}\mbox{C}}$
  of the residual nucleus $^{14}\mbox{C}$ 
 is taken into account by performing in (\ref{op-equation}) the transformation 
 \cite{Gui91} ($i \neq j \in 1,2$)
\begin{equation}\label{trafo1}
\vec{p}^{\,\,0}_i \rightarrow \vec{q}^{\,\,0}_i = 
\frac{1}{m_{^{16}\mbox{O}}}\left[
(m_p + m_{^{14}\mbox{C}} )
  \vec{p}^{\,\,0}_i - m_p (\vec{p}^{\,\,0}_j + \vec{p}_B),
\right]
\end{equation}
where $m_p (m_{^{16}\mbox{O}})$ denotes the mass of the outgoing proton 
 ($^{16}\mbox{O}$) and $\vec{p}_B$ the recoil momentum of 
 the residual nucleus $^{14}\mbox{C}$. Moreover,   a semirelativistic 
generalization  of (\ref{op-equation}) has been used  
  as discussed in \cite{Nad81}.

In all previous work, the mutual NN-interaction 
$V^{NN}$   between   the two outgoing protons (NN-FSI) was neglected.
 In the present study,
  this approximation is dropped by incorporating for the first time 
  the corresponding complete NN-scattering amplitude $T^{NN}$
 ($z= \frac{(\vec{p}^{\,0}_1)^2}{2m_p} + \frac{(\vec{p}^{\,0}_2)^2}{2m_p} 
+ i \epsilon$)
\begin{equation}\label{tnn}
T^{NN}(z) = V^{NN} + V^{NN} G_0(z) T^{NN}(z), 
\end{equation}
with
\begin{equation}\label{g0}
G_0(z) = \frac{1}{z-H_0(1)-H_0(2)},
\end{equation} 
  up to the first order in the final state, as it has been depicted in 
  diagram (b) of fig.\  \ref{fig_mechanism}. 
For the sake of  simplicity, multiscattering processes like those
described by diagrams (c) and (d) of fig.\   \ref{fig_mechanism}
 are still neglected. The final state  $\ket{\Psi_f}$ in (\ref{eq1})
 is therefore  given in our approach by
\begin{equation}\label{fsi-state}
 \ket{\Psi_f} = \ket{\phi^{OP}(\vec{q}^{\,\,0}_1)}\,
\ket{\phi^{OP}(\vec{q}^{\,\,0}_2)} + G_0(z) T^{NN}(z) 
\ket{\vec{p}^{\,\,0}_1}\,\ket{\vec{p}^{\,\,0}_2},
\end{equation}
where $\ket{\vec{p}^{\,\,0}_i}$ denotes a plane wave state of the 
 proton $i$ with  momentum $\vec{p}^{\,\,0}_i$.  Within this treatment of FSI,
we are still far away from having solved 
  the complete three-body problem of the final state. Nevertheless,
 we are  able to obtain 
a first reliable estimate  of  the relevance of  NN-FSI
in various kinematical situations of two-proton knockout.  

 In our explicit evaluation, we have used in (\ref{tnn})
 as NN-potential  $V^{NN}$  the Bonn OBEPQ-A potential \cite{Mac89}
 which has
 also been used for the calculation of the defect functions in the 
 initial state. Due to the nonlocality of this potential, the
  term $G_0(z) T^{NN}(z) 
\ket{\vec{p}^{\,\,0}_1}\,\ket{\vec{p}^{\,\,0}_2}$ in (\ref{fsi-state})
 is explicitly evaluated in momentum space. The initial state 
 $\ket{\Psi_i}$  and the  nuclear current $\hat{J}^{\mu}(r)$ in (\ref{eq1})
 are however calculated  in configuration space, so that finally
 an appropriate Fourier transformation
 of the NN-FSI from momentum to configuration space had to be  performed. 
 Moreover, we would like to mention that a  usual partial wave decomposition 
 \cite{MaH87,ScA01} of the 
 NN-interaction $V^{NN}$ has been adopted taking into account 
 all isospin 1 partial NN-waves up to an orbital angular momentum of 3,
  i.e.\  the $^1S_0, ^3P_0, ^3P_1, ^3P_2, ^1D_2, ^3F_2$, $^3F_3$
 and $^3F_4$ contributions.
  It has been checked numerically that the contribution of $G$ and $H$ NN-waves
  is negligible, at least   for the kinematics considered
  here.

\section{Results}
\label{res}
In this section, we discuss the influence of NN-FSI for two different
types of kinematics which have already been under experimental 
 investigation \cite{Ros00,Ond97,Ond98}.
 We call $E_0$ the incident electron energy and $\theta_e$ the electron 
 scattering angle in the laboratory frame.
 The energy and momentum transfer is denoted, as usual, 
 as $\omega$ and $q$, respectively. The 
 angles between the momentum transfer $\vec{q}$ and the momenta 
 $\vec{p}^{\,\,0}_1$ and $\vec{p}^{\,\,0}_2$ of the outgoing protons are 
 called $\gamma_1$ and $\gamma_2$.

Concerning the different approximations for the final state, 
 we denote as PW the plane-wave approximation, where FSI are completely
 neglected, and  as DW  the treatment of \cite{Gui97}, where only the 
 optical potential $V^{OP}$  is taken into 
 account. In the approach PW-NN we consider the alternative case where 
only $V^{NN}$, but not  $V^{OP}$, is included. The corresponding
  final states in these different approximations are given by
\begin{eqnarray}\label{fsi-approx}
 \ket{\Psi_f}^{PW} &=&
 \ket{\vec{q}^{\,\,0}_1}\,\ket{\vec{q}^{\,\,0}_2}\,\, , \\
 \ket{\Psi_f}^{DW} &=&
 \ket{\phi^{OP}(\vec{q}^{\,\,0}_1)}\,\ket{\phi^{OP}(\vec{q}^{\,\,0}_2)}
\,\, , \\
\ket{\Psi_f}^{PW-NN} &=& \ket{\vec{q}^{\,\,0}_1}\,\ket{\vec{q}^{\,\,0}_2}
+  G_0(z) T^{NN}(z) 
 \ket{\vec{p}^{\,\,0}_1}\,\ket{\vec{p}^{\,\,0}_2}\,\,. 
\end{eqnarray}
Our full approach in (\ref{fsi-state}) is denoted as
DW-NN.

\begin{figure}
\centerline{\psfig{figure=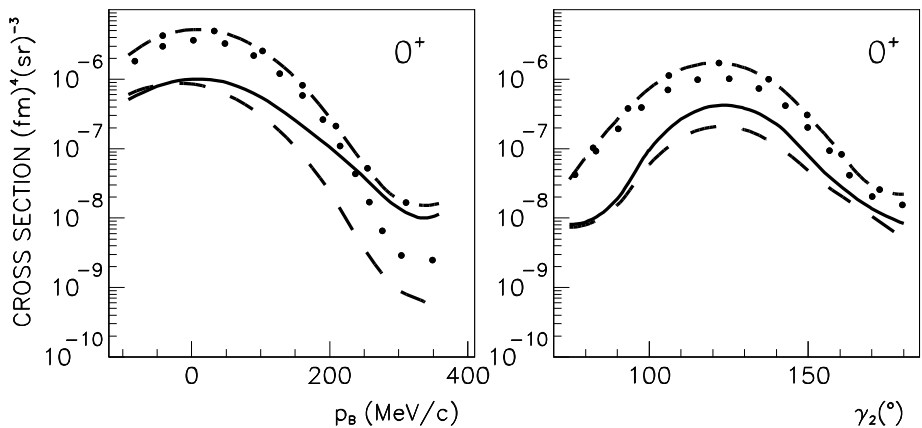,width=14cm,angle=0}}
\vspace{0.5cm}
\caption{The differential cross section  in the $^{16}\mbox{O}(e,e' pp)$ 
 reaction
to the $0^+$ ground state of $^{14}\mbox{C}$
 for the two kinematics discussed  in the text.
 Used notation for the different calculations:
 PW (dotted), PW-NN (dash-dotted), DW (dashed), DW-NN (solid).}
\label{fig2}
\end{figure}

The results of these different approaches on the cross section of the 
$^{16}\mbox{O}(e,e' pp)$ 
 reaction for the transition to the $0^+$ ground state of 
$^{14}\mbox{C}$ 
 are shown in fig.\  \ref{fig2}. In the left panel the superparallel
kinematics of a Mainz experiment \cite{Ros00} is considered,  with  $E_0= 855$ 
MeV, $\theta_e = 18^{\circ}$, $\omega=215$ MeV, $q=316$ MeV/$c$, 
$\gamma_1= 0^{\circ}$, and $\gamma_2 = 180^{\circ}$. In the right panel 
an alternative kinematical setup, which has been  included in a NIKHEF 
experiment \cite{Ond97,Ond98},  is investigated, with  $E_0= 584$ MeV, 
$\theta_e = 26.5^{\circ}$, $\omega=210$ MeV and  $q=300$ MeV/$c$. The angle 
$\gamma_1$ is $30^{\circ}$, on the opposite side of the outgoing electron with 
respect to $\vec{q}$. The kinetic energy of proton 1 is fixed to 
$T_1 = 137$ MeV.

By changing  the kinetic energy of the outgoing protons 
in the superparallel kinematics and the angle $\gamma_2$ in the NIKHEF setup,
 we are able to  explore different values  of  the recoil momentum of the 
 residual nucleus $p_B \equiv|\vec{p}_B|$. Positive (negative) values of 
$p_B$ in the left panel refer to situations where 
$\vec{p}_B$ is parallel (antiparallel) to the momentum transfer. 
It is well known and can be clearly seen in fig.\  \ref{fig2} that the 
inclusion of the optical potential leads to an overall and substantial 
reduction of the cross section in both kinematical setups (consider the 
difference between the PW and DW results). On the other hand,
 our calculations give a considerable enhancement for medium and large 
 recoil momenta in the superparallel kinematics if  NN-FSI are  taken into 
 account (see the difference between  PW and PW-NN,  and between DW and DW-NN).
 The effect of NN-FSI amounts to about one order of magnitude
 enhancement  at $p_B  = 300$ 
MeV/$c$. Thus even at  back-to-back kinematics the 
mutual interaction of the two outgoing protons cannot be neglected. In the 
NIKHEF kinematics, the effect of NN-FSI is also sizeable, although not as 
strong as in the superparallel kinematics. Moreover, whereas in the 
superparallel kinematics the relative effect  of NN-FSI  increases for 
decreasing cross section,  in the NIKHEF kinematics NN-FSI is maximal when 
also the cross section is maximal, i.e. for $\gamma_2 \approx 120^{\circ}$, 
which corresponds to $\vec{p}_B \approx 0$ MeV/$c$.  
This result clearly shows that the role of NN-FSI is strongly dependent 
on the kinematics and no general statement can be drawn with respect to its 
relevance.

\begin{figure}
\centerline{\psfig{figure=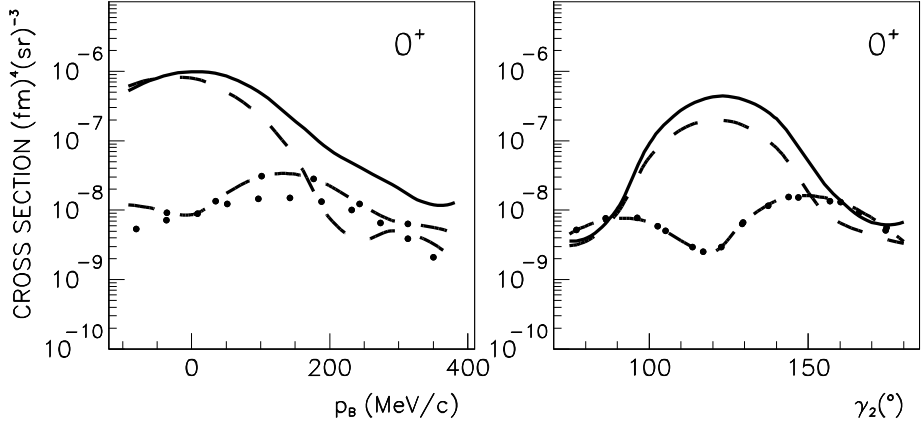,width=14cm,angle=0}}
\vspace{0.5cm}
\caption{The differential cross section  in the $^{16}\mbox{O}(e,e' pp)$ 
 reaction
to the $0^+$ ground state of $^{14}\mbox{C}$ 
 in the same two kinematics as in fig.\  
\ref{fig2}. The dashed (solid) curve shows the separate contribution  of the  
$^1S_0$ relative partial wave  in a  DW (DW-NN) calculation. The dotted 
(dash-dotted) curve shows the separate contribution  of the  $^3P_1$ relative 
partial wave  in a  DW (DW-NN) calculation.}
\label{fig3}
\end{figure}

As is known from previous work \cite{Gui97}, in the 
 $^{16}\mbox{O}(e,e' pp)$ 
reaction the transition to the $0^+$ ground state of $^{14}\mbox{C}$
   is governed
dominantly  by the $^1S_0$ partial wave in 
 the initial relative state of the two protons. A 
sizeable contribution arises  moreover from  the $^3P_1$ state.
 The relative importance of NN-FSI on these two partial waves is presented
 in fig.\  \ref{fig3} for the kinematical setups already considered in fig.\ 
 \ref{fig2}. In both kinematics the effect of NN-FSI is more important on
 the $^1S_0$ initial state. The effect on this single state gives in practice 
 almost the full contribution of NN-FSI. For the  $^3P_1$ initial state only a 
 negligible effect is given in the NIKHEF kinematics. The effect is
 somewhat larger in the superparallel kinematics, but also here it is completely
 overwhelmed in the final result by the dominant contribution of the $^1S_0$ 
 state. 
 
\begin{figure}
\centerline{\psfig{figure=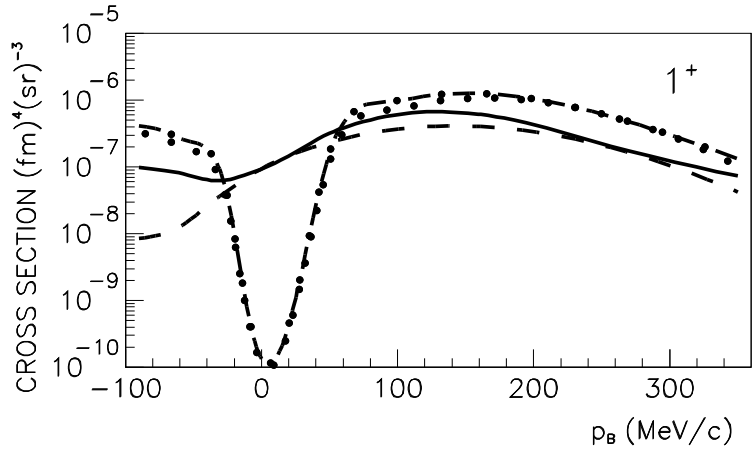,width=12cm,angle=0}}
\vspace{0.5cm}
\caption{The differential cross section  in the $^{16}\mbox{O}(e,e' pp)$
  reaction
to the $1^+$  state of $^{14}\mbox{C}$
  for the superparallel kinematics discussed
 in the text. Line notation as in fig.\  \ref{fig2}.
}
\label{fig4}
\end{figure}

 The role of NN-FSI on the $^3P$ initial relative states is of specifical
 relevance for the transition to the $1^+$ excited state of $^{14}\mbox{C}$,  
 where only $^3P$ components are present and the $^1S_0$ relative partial 
 wave cannot contribute. The results for this transition in the superparallel
 kinematics are depicted in fig.\  \ref{fig4}. A negligible effect is due to
 NN-FSI in the PW-NN approach. A significant enhancement is obtained in the 
 DW-NN calculation, especially for negative values of $p_B$, 
 where the cross section has a minimum.
     
Summarizing, we have studied the  importance  of the mutual final state 
interaction of the two emitted protons  in the $^{16}\mbox{O}(e,e' pp)$ 
 reaction 
within a perturbative treatment. Our results indicate that NN-FSI can 
 be important in particular situations and in general cannot be neglected.
 It has been clearly shown that the role of NN-FSI depends on  kinematics and 
 on the final state in the excitation spectrum of $^{14}\mbox{C}$.
  Therefore, one may 
 hope that it is possible to find specific kinematical situations where the  
 effects  of  FSI  are as small  as possible, in order to achieve the most 
 direct access to SRC in complex nuclei. This requires a systematic 
 study of the role of FSI for different kinematics which is presently under 
 consideration. Moreover, due to our numerical  results, the full 
 three-body problem  of the final state has to be tackled
 in forthcoming studies.  In that 
 context, special emphasis has to be devoted to a more consistent  treatment 
 of the initial and the final state in our approach. 

\vspace{0.2cm} 
\centerline{{\bf Acknowledgements}}
\vspace{0.2cm}
 This work has been supported by the
 the Istituto Na\-zionale di Fisica Nucleare and by the
 Deutsche For\-schungs\-gemein\-schaft (SFB 443).
 M. Schwamb  would like to thank  the Dipartimento di Fisica
 Nucleare e Teorica of the University of Pavia for the warm hospitality
 during his stays in Pavia. Fruitful discussions with H. Arenh\"ovel  are
 gratefully acknowledged.

\end{document}